\documentclass{PoS}

\usepackage{graphics,graphicx}% Include figure files
\usepackage{bm}% bold math
\usepackage{epsfig}
\usepackage{mathtools}
\usepackage{epstopdf}
\usepackage{slashed}
\usepackage{stackrel}
\usepackage[normalem]{ulem}

\title{Aspects of heavy flavor jet physics in heavy ion collisions}

\ShortTitle{Heavy flavor jets}

\author{\speaker{Ivan Vitev}%\thanks{A footnote may follow.}
\\
  Los Alamos National Laboratory, Theoretical Division, Mail Stop B283,
       Los Alamos, NM 87544 \\
        E-mail: \email{ivitev@lanl.gov}}

\abstract{In these proceedings I discuss several recent developments in the physics of heavy flavor jets in
heavy ion collisions. i) The dijet mass modification in nucleus-nucleus reactions has been proposed as a new observable with enhanced sensitivity  to parton energy loss in nuclear matter. It also enables more precise studies of heavy quark mass effects on parton shower  formation.  ii)  Computational techniques from soft-collinear effective theory have allowed us to bridge the gap between high energy and heavy ion QCD phenomenology. I show the first application of the semi-inclusive jet function formalism to heavy flavor jet production in proton-nucleus and nucleus-nucleus collisions at the LHC. iii)  Last but not least, central to the theoretical calculations of heavy flavor jets 
is the accurate theoretical description of in-medium parton showers. A formalism to compute branching processes in nuclear matter to any desired order in opacity has been developed and illustrative numerical results are presented.     }

\FullConference{13th International Workshop in High pT Physics in the RHIC and LHC Era (High-pT2019)\\
		19-22 March 2019\\
		Knoxville, Tennessee, USA}

\begin{document}

\section{Introduction}

Theoretical and experimental studies of heavy flavor are central to high-energy nuclear physics. They provide 
new avenues to explore Quantum Chromodynamics and new diagnostics of the
transport properties of  nuclear matter matter ~\cite{Andronic:2015wma}.  In these proceedings I will focus 
on open heavy flavor and jets that contain charm ($c$) and beauty ($b$) quarks in particular.  
Theoretical investigation of  heavy-flavor tagged jet production in heavy ion collisions has been somewhat 
limited~\cite{Huang:2013vaa,Senzel:2016qau,Li:2017wwc,Dai:2018mhw},  even though experimental effort  
in this direction is ramping up~\cite{Chatrchyan:2013exa,Sirunyan:2018jju,hassan:hal-01846896,Adare:2015kwa}.  At transverse  momenta  $p_T \geq 100$~GeV the suppression of $b$-jets in nucleus-nucleus (A+A) reactions relative to the proton-proton (p+p) baseline was found to be similar to the one for inclusive jets.  In the quest to identify
observables that are sensitive to the heavy heavy quark mass, groomed soft-dropped momentum sharing  
distributions in heavy ion collisions show promise at transverse momenta $\leq  50$~GeV~\cite{Li:2017wwc}.  An alternative  strategy is to exploit the fact that multi-jet events are abundant in high energy hadronic collisions  and
devise observables, such as dijet mass,  that amplify the effects of quenching~\cite{Kang:2018wrs}.

Until recently, calculations of heavy flavor jet observables were limited to the traditional charm and beauty quark energy loss approach, see for example~\cite{Djordjevic:2003zk}. It is now possible to evaluate $b$-jet and $c$-jet cross sections using the technique of semi-inclusive jet functions~\cite{Li:2018xuv}.  This theoretical development was enabled by the recent derivation of the semi-inclusive jet functions  for heavy flavor jets in the vacuum~\cite{Dai:2018ywt},  and the
in-medium splitting kernels for heavy flavor to first order in opacity~\cite{Kang:2016ofv}. Phenomenological results 
give a good description of the LHC experimental measurements with improved control over theoretical uncertainties. 
Effective field  theories for heavy jet substructure are also being developed, first in p+p collisions~\cite{Lee:2019lge}.
This serves as a motivation to  revisit  and further improve the  Altarelli-Parisi splitting functions in nuclear matter that are the key ingredients in  high-precision calculations in the theory of strong interactions and in Monte-Carlo event generators for heavy ion physics~\cite{Sievert:2019cwq}. The evaluation of correlation effects due to coherent multiple scattering on parton branching processes is discussed.

These developments are outlined below in more detail: the dijet mass calculations are shown in Section~2. Section~3 summarizes the semi-inclusive jet function approach to $b$- and $c$-jets. Parton branching calculations 
beyond the soft gluon emission limit to higher order in opacity follow in Section~4.  Conclusions are given in Section~5.

\section{Back-to-back dijets and dijet mass modification}

\begin{figure}[t]
    \centering
    \includegraphics[width=0.49\textwidth]{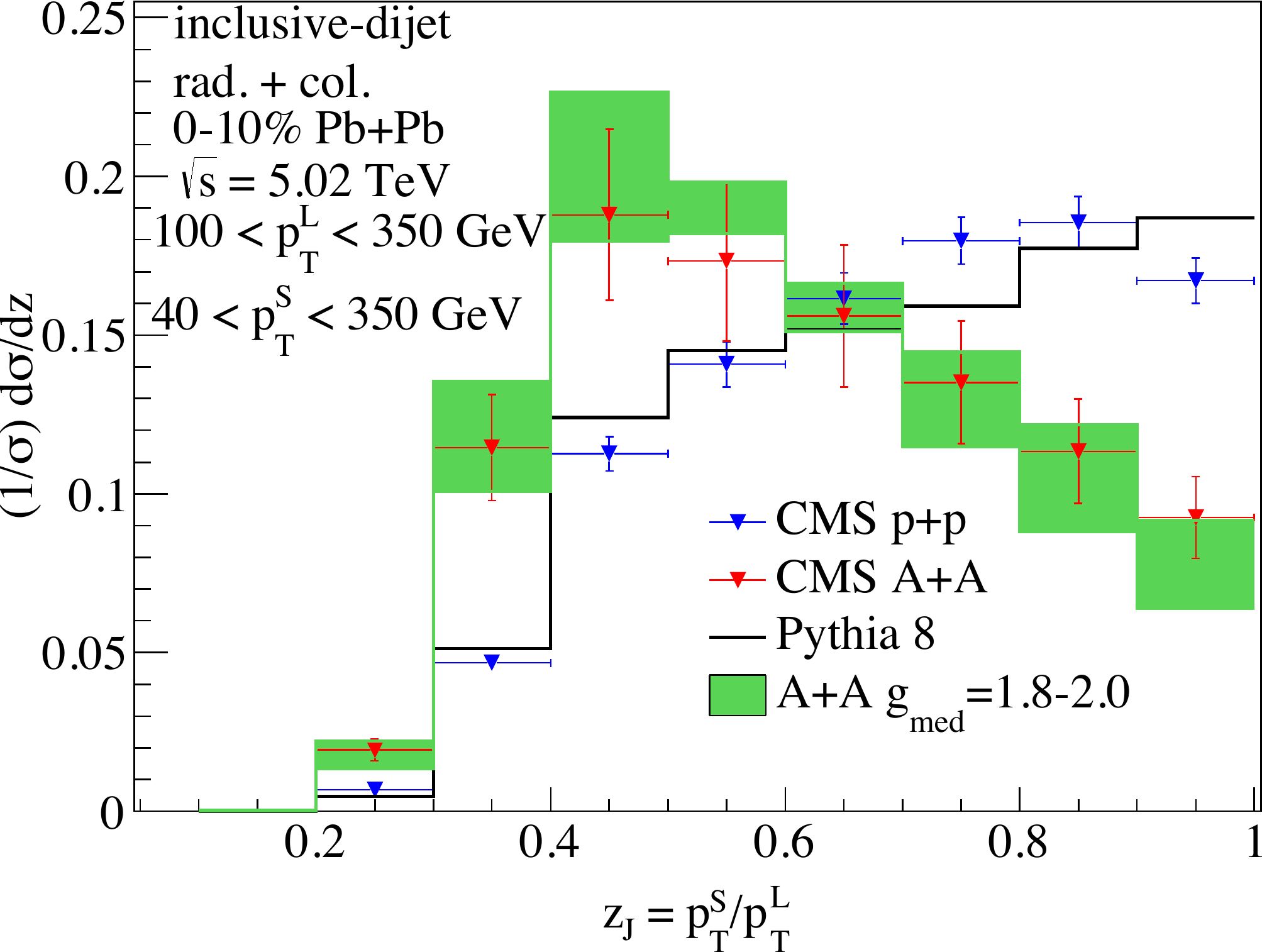}  \hspace*{0.1cm}
       \includegraphics[width=0.49\textwidth]{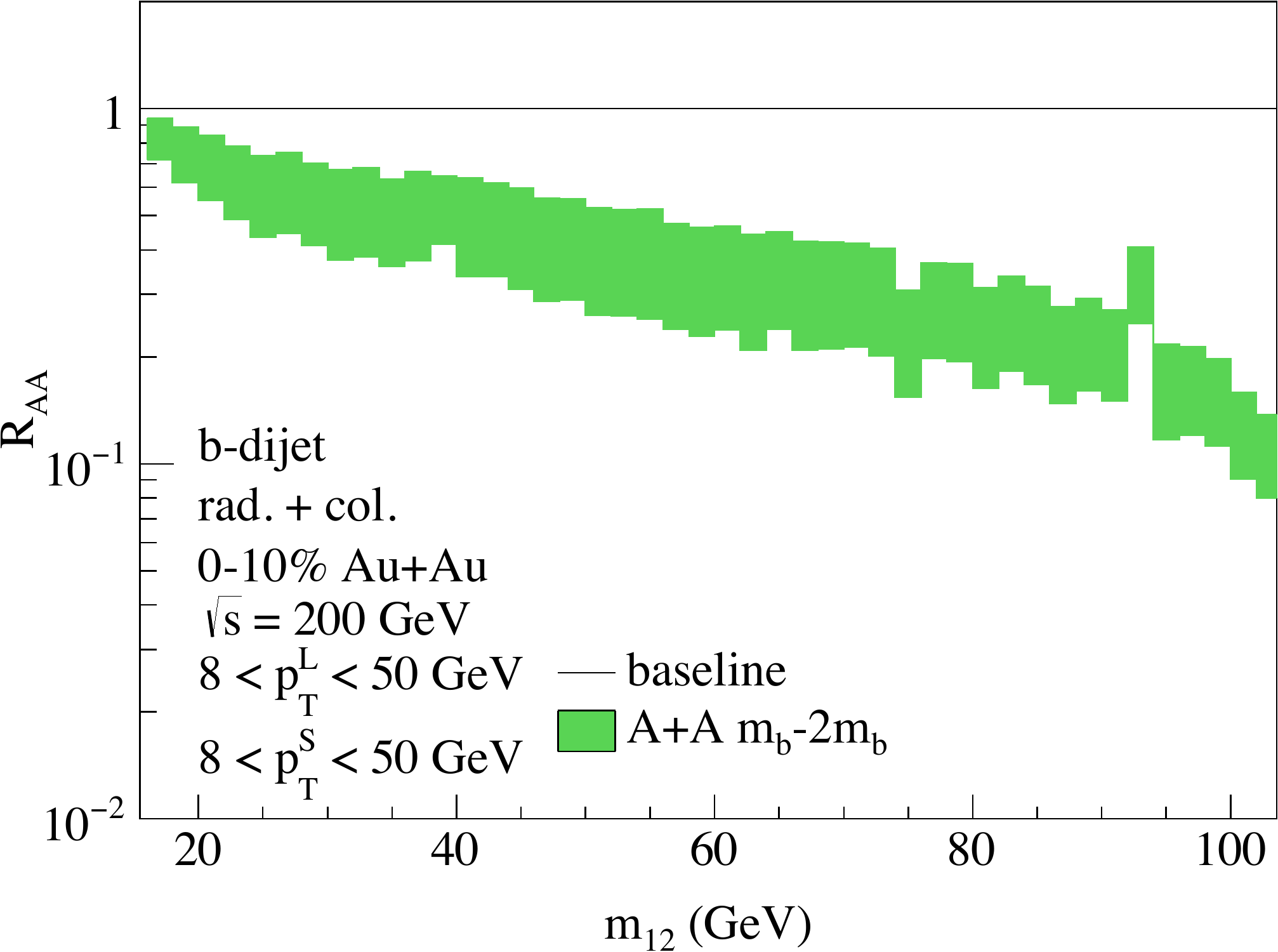}
    \vspace{-0.5cm}
    \caption{Left: the dijet imbalance $z_J$ distributions for inclusive jets  at $\sqrt{s_{NN}} = 5.02$~TeV  are compared to CMS collaboration data~\cite{Sirunyan:2018jju}. The band corresponds to a range of coupling strengths between the jet and the medium: $g_{\rm med}=1.8-2.0$. Right:  nuclear modification factor plotted as a function of dijet invariant mass $m_{12}$ for $b$-tagged (right) dijet production in Au+Au collisions at $\sqrt{s_{NN}} = 200$ GeV at sPHENIX. We fix $g_{\rm med}=2.0$, and the band corresponds to a range of masses of the propagating system between $m_b$ and $2m_b$. Figures are reproduced from Ref.~\cite{Kang:2018wrs}.}
    \label{fig:dijets}
\end{figure}

Back-to-back light and heavy flavor dijet measurements are important for the accurate study  of jet production and propagation in a dense QCD medium. They can test the path length, color charge, and mass dependence of quark and gluon energy loss in the quark-gluon plasma (QGP) produced in ultra-relativistic nucleus collisions.  
Traditional studies of dijets  in heavy ion physics have focused on observables such as the dijet momentum imbalance shift -  the normalized distribution of the imbalance variable $z_J = p_{2T}/p_{1T}$.   In the absence of a QGP one expects  that the transverse momenta of the two jets are approximately balanced. On the other hand, in heavy ion collisions one of the jets may lose more energy than the other, resulting in a downshift of the peak in $z_J$ distribution because of strong in-medium interactions. This is shown in the left panel of Figure~\ref{fig:dijets}, and while the shapes in A+A and p+p reactions are easy to differentiate,  the changes in the average
momentum imbalance $\langle z_J \rangle $ are subtle, on the order if 10\%.

As an example of observable where jet quenching effects ``add'' rather than ``subtract'' we look at the
 dijet invariant mass $m_{12}^2 = (p_1 + p_2)^2$ . It can be written in terms of the jets' transverse momentum and rapidity as follows
\begin{equation}
m_{12}^2 = m_1^2 + m_2^2 + 2\left[m_{1T}m_{2T}\mathrm{cosh}(\Delta \eta) - p_{1T}p_{2T}\mathrm{cos}(\Delta \phi)\right],
\end{equation}
where $m_i^2 = p_i^2$ and $m_{iT} = \sqrt{m_i^2 + p_{iT}^2}$ are the invariant mass squared and the transverse mass for one of the jets; $\Delta \eta$ and $\Delta \phi $ are the differences in the rapidities and the azimuthal angles,
respectively.  We can evaluate the dijet mass distribution from the double differential dijet cross section as follows
\begin{equation}
\frac{d\sigma}{dm_{12}} = \int dp_{1T} dp_{2T} \frac{d\sigma}{dp_{1T}dp_{2T}} \delta\left(m_{12} - \sqrt{\langle m_1^2\rangle + \langle m_2^2\rangle + 2p_{1T}p_{2T}\langle\mathrm{cosh(\Delta \eta)} - \mathrm{cos}(\Delta \phi)\rangle} \right),
\label{eq:mass}
\end{equation}

We find that the suppression of the dijet invariant mass can be as large as a factor of 10 at both RHIC and the LHC. For reference,  the typical suppression of inclusive  jets is typically a factor of 2.  The nuclear modification is characterized by enhanced sensitivity to the coupling between the jet and the medium, for example a  10\% variation in $g$ can lead to more than a factor of 2 difference  in the suppression, denoted $R_{AA}$.  Heavy quark mass effects are also more pronounced.  One example is shown in the right panel of  Figure~\ref{fig:dijets}, where the reduction in $R_{AA}$ toward smaller dijet masses  is purely driven by  the  large heavy quark mass $m_b$.  The steeply falling spectra at RHIC energies facilitate such differentiation, but dijet masses can also be very effectively studied at the LHC.

\section{Inclusive $b$-jet production from semi-inclusive jet functions}

Inclusive jet production in  both  proton-proton (p+p) and heavy ion  (p+A, A+A) collisions is a multiscale problem, suited to effective field theory treatment. The differential jet cross section versus   $p_T$ and  $\eta$ in hadronic collisions can be expressed as convolution of the parton distribution functions (PDFs), the hard part, and the semi-inclusive jet functions (SiJFs):
%\begin{widetext}
\begin{align}\label{eq:main}
    \frac{d\sigma_{pp\to J +X}}{dp_T d\eta} =& \frac{2p_T}{s}\sum_{a,b,c} \int_{x_a^{\rm min}}^1 \frac{dx_a}{x_a} f_a(x_a,\mu) \int_{x_b^{\rm min}}^1 \frac{dx_b}{x_b} f_b(x_b,\mu)
    \nonumber \\ &\times
    \int_{z_{\rm min}}^1 
     \frac{dz_c}{z_c^2}
    \frac{d\hat{\sigma}_{ab\to c}(\hat{s}, p_T/z_c, \hat{\eta}, \mu)}{dv dz} J_{ J/c}(z_c,  w_J \tan(R^\prime/2), m_Q, \mu)~. 
\end{align}
%\end{widetext}
Here $f_{a,b}$ are the  PDFs, $d\hat{\sigma}_{ab\to c}(\hat{s}, p_T, \hat{\eta}, \mu)/dv dz$ is the hard function for the sub-process $a b \to c $, and  $J_{ J/c}$ is the jet function.  It describes the probability of a parton $c$ with transverse momentum $p_T/z_c$ to fragment into a jet $J$ with $p_T$.  The $\ln R$ resummation for  jet production can be achieved by evolving  the  SiJFs from the jet scale $\mu_J$ to the factorization scale $\mu$. The renormalization group  equations are the usual time-like DGLAP evolution equations.

\begin{figure}[t]
    \centering
    \includegraphics[width=0.49\textwidth]{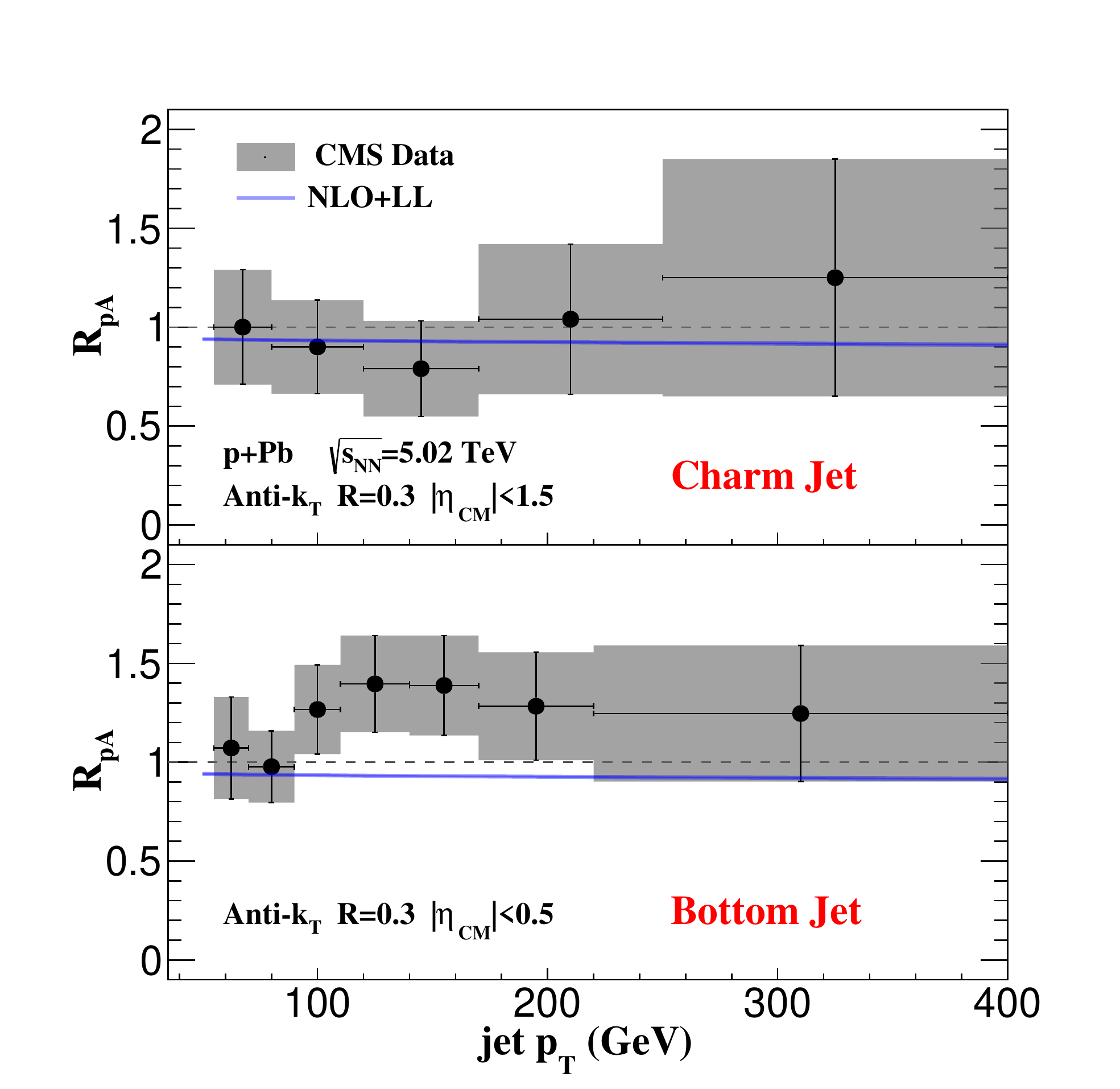}  \hspace*{0.1cm}
       \includegraphics[width=0.49\textwidth]{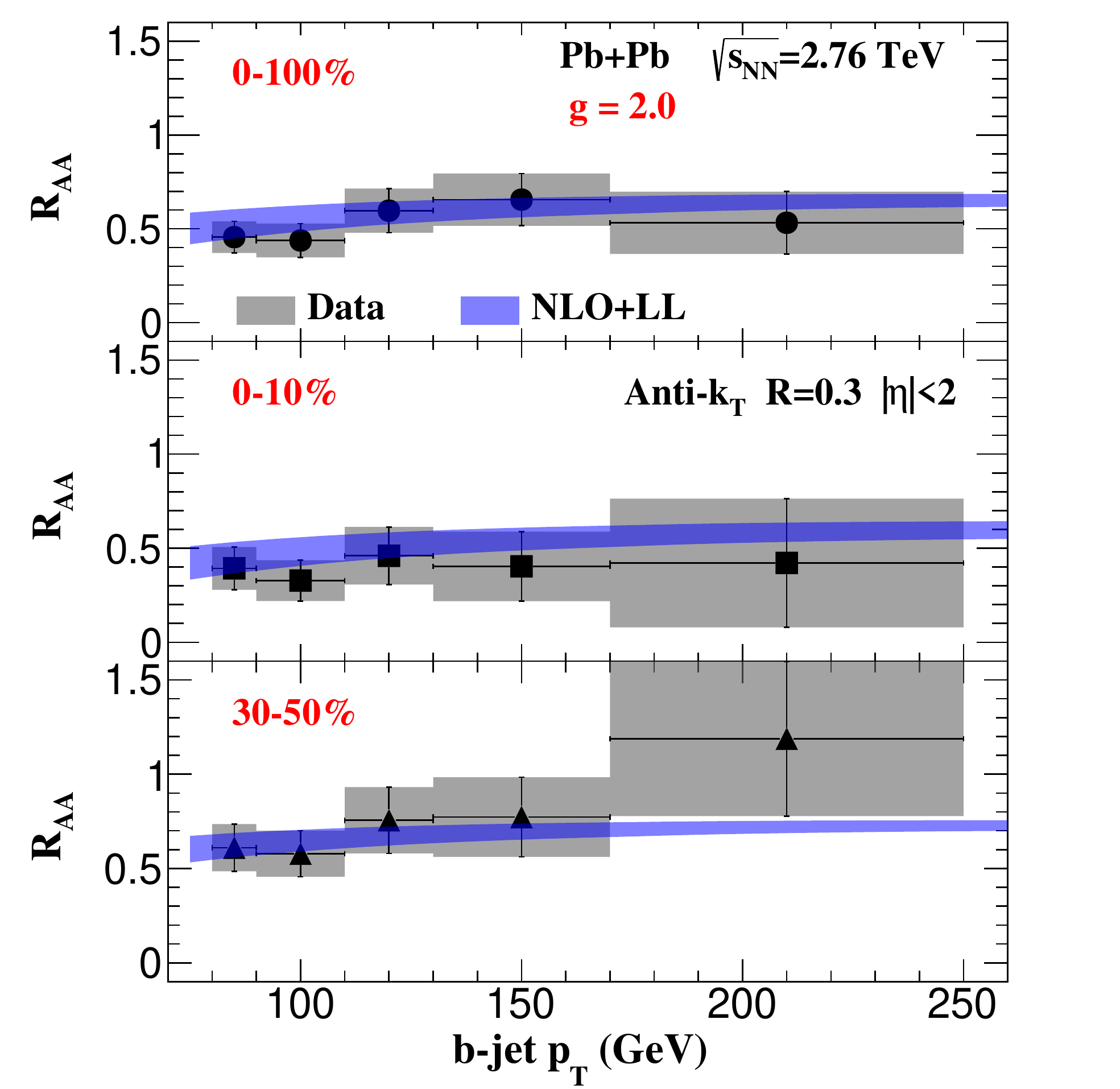}
    \vspace{-0.5cm}
    \caption{Left: comparison of predicted heavy flavor jet cross section $R_{\rm pA}$  in proton-nucleus collisions to CMS measurements~\cite{Sirunyan:2016fcs,Khachatryan:2015sva} of $c$-jets (top) and $b$-jets (bottom)  at $\sqrt{s_{\rm NN}}=5.02$~TeV.   Right: the nuclear modification factor  $R_{AA}$  of  $b$-jets    for different centrality classes (0-100\%, 0-10\% and 30-50\% ), as indicated  in the legend.  Data is from CMS measurements~\cite{Chatrchyan:2013exa}.  Figures are reproduced from Ref.~\cite{Li:2018xuv}. }
    \label{fig:RpA}
\end{figure}

High energy jet production in p+A reactions places constraints on cold nuclear matter (CNM) effects.  Within the statistical and systematic uncertainties many experimental measurements are consistent with a range of possibilities - from no nuclear effects to $\pm10\%$ cross section modification. The model described here  takes into  
account the initial-state CNM energy loss. It is compared to data from 5.02~TeV p+Pb collisions~\cite{Sirunyan:2016fcs,Khachatryan:2015sva} in  the left panel of Figure~\ref{fig:RpA}.  In nucleus-nucleus collisions the semi-inclusive  jet functions receive corrections from in-medium energy dissipation and QGP-induced branching. 
For example, the NLO medium correction to the $Q\to J_{Q}$ SiJF is  
\begin{align}
    J^{\text{med},(1)}_{J_Q/Q}(z, p_T R,m, \mu) =&  \int_{z(1-z) p_T R }^{\mu} d q_\perp P^{\rm med}_{QQ}(z, m, q_\perp) 
   - \delta(1-z) \int_0^1 dx  \int_{x(1-x)  p_T R}^{\mu} d q_\perp P^{\rm med}_{QQ}(x, m, q_\perp)   
    \nonumber \\
    = & \left[\int_{z(1-z)  p_T R}^{\mu} d q_\perp P^{\rm med}_{QQ}(z, m, q_\perp) \right]_+~.
\end{align}
In the above equation, the terms correspond to the contribution with a radiation outside of the jet cone and the combination of the real radiation inside the jet cone and the virtual loop corrections.
$P^{\rm med}_{QQ}(z, m, q_\perp)$ is the medium-induced heavy quark splitting kernel.

Numerical calculations of $b$-jet suppression are compared to data~\cite{Chatrchyan:2013exa} 
 in the right panel of Figure~\ref{fig:RpA}.  $R_{\rm AA}$ decreases (indicating larger suppression) with increasing collision centrality. The attenuation factor is less dependent on the centrality when compared to the  light jet modification. The predictions agree very well with the data for both  the inclusive cross sections and the nuclear modification factors.

\section{In-medium  splitting functions to any order in opacity}

Advances in the theoretical understanding and  experimental measurements of reconstructed jets and heavy flavor necessitate  more precise control over  in-medium branching processes.  An important step in this direction is to obtain in-medium splitting kernels beyond the soft gluon  approximation and to higher orders in opacity - the correlation between multiple scattering centers in the  nuclear medium. %
In Ref.~\cite{Sievert:2018imd}  a lightcone wavefunction techniques were combined with recurrence relations to 
allow for the calculation medium-induced splitting processes  to an arbitrary order in opacity.  
One should note that it was argued that full numerical evaluation of such processes is necessary, as corrections to simple analytic limits are large~\cite{Feal:2018sml}.

\begin{figure}[tb]
	\begin{center}
		\includegraphics[width=0.46 \textwidth]{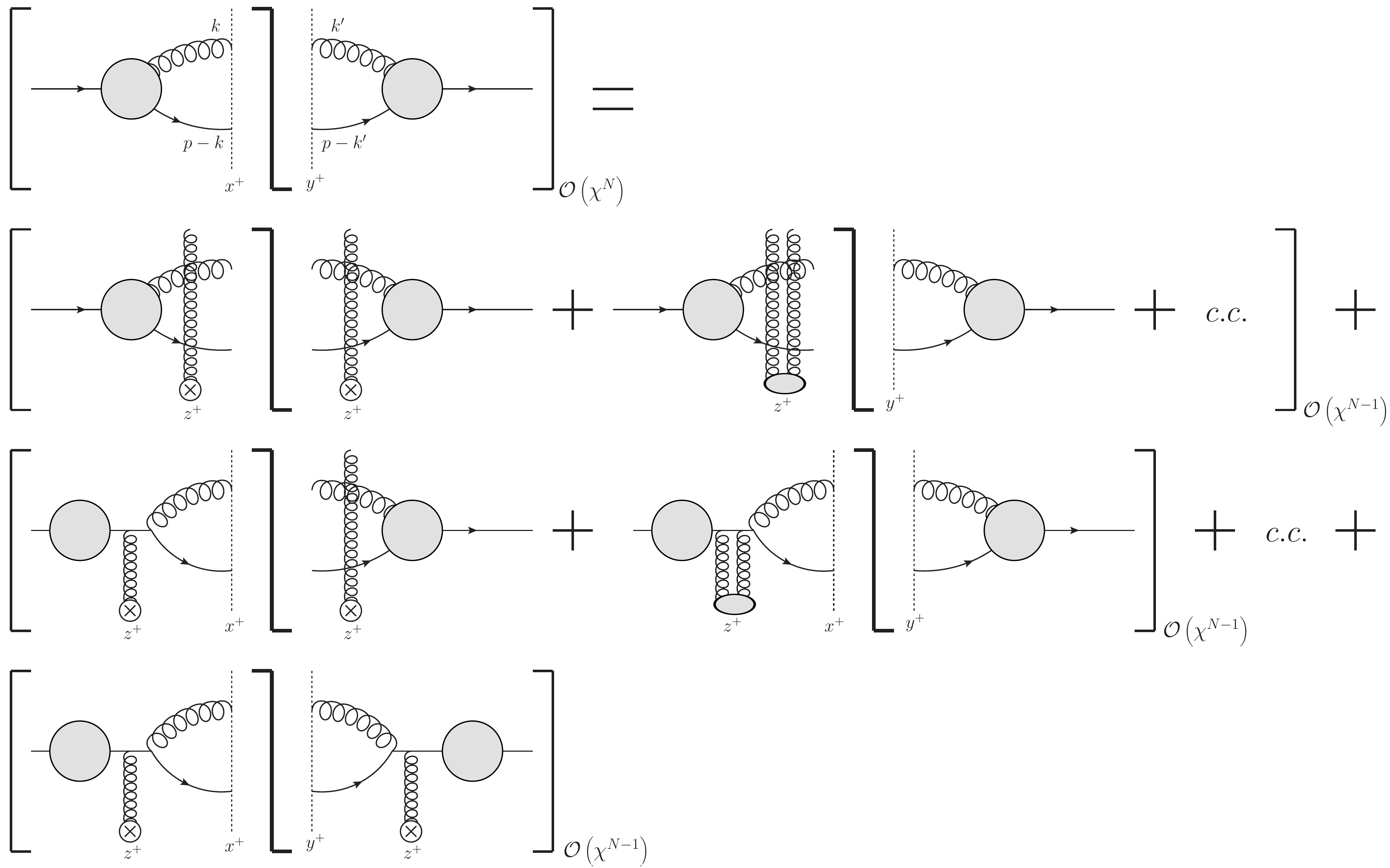}  \hspace*{1cm}
			\includegraphics[width= 0.46\textwidth]{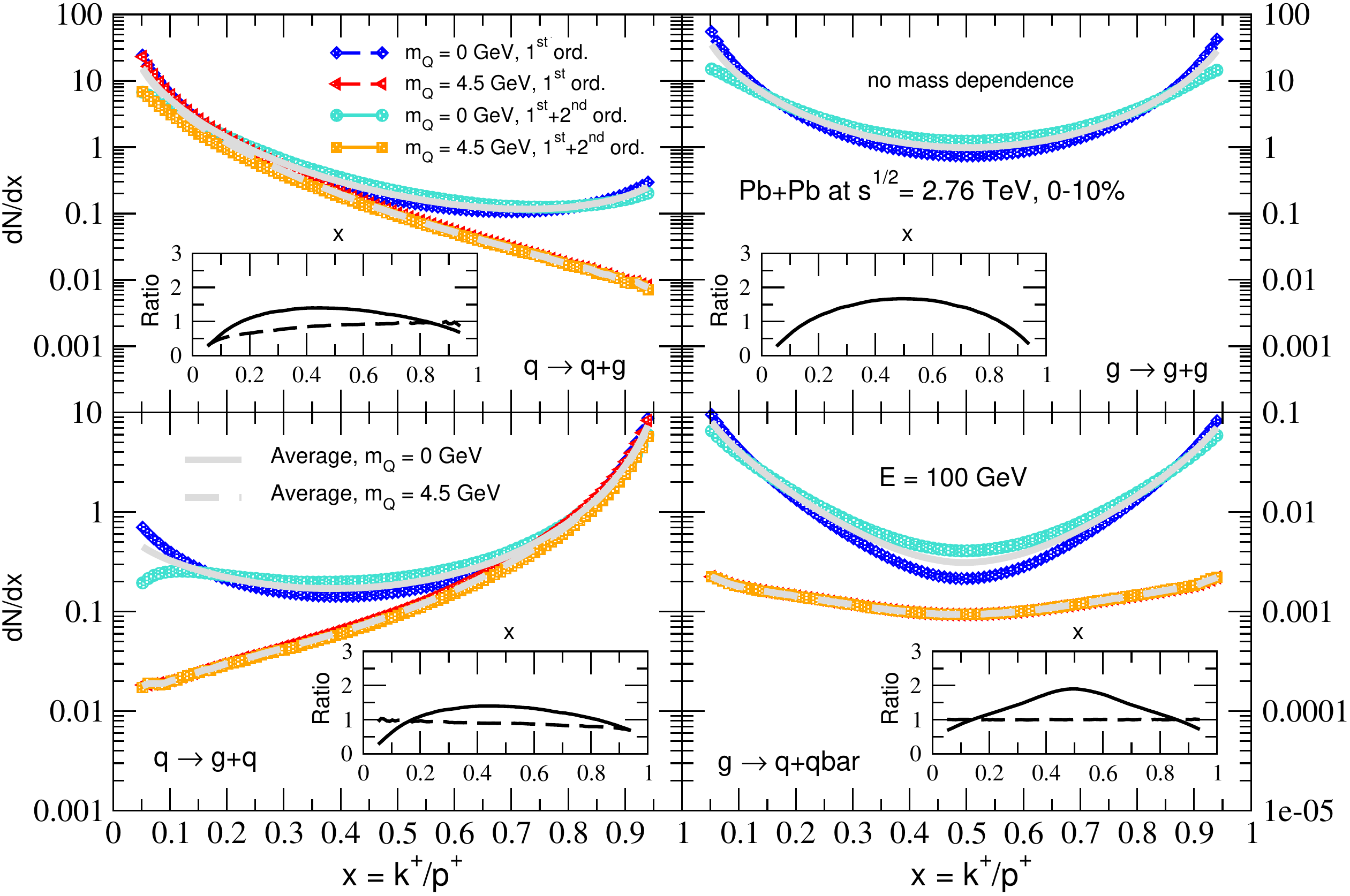} 
		\caption{Left: diagrammatic representation of the recursion relation  for the Final/Final sector of the $q \rightarrow q g$ branching channel.   Right: the one-dimensional differential splitting functions $\frac{dN}{dx}$ for a $100 \, \mathrm{GeV}$ jet as a function of the splitting fraction $x$.   Both light and heavy quark results are shown where relevant.  Insets: The ratio of the $(1^{\mathrm{st}} + 2^{\mathrm{nd}}) / 1^{\mathrm{st}}$ order results.
Figures are reproduced from Ref.~\cite{Sievert:2019cwq}		}
		\label{f:ReactFF}
	\end{center}
\end{figure}

The original work, which focused on the single $q \rightarrow qg$ chanel, has now ben extended  to include all
4 lowest order branching processes in the medium ($q \rightarrow qg$, $g\rightarrow gg$, $q \rightarrow g q$, $g \rightarrow q\bar{q}$). The type of single-Born and double-Born interactions between the propagating partonic system and the
nuclear medium that we consider are shown in the left panel of Figure~\ref{f:ReactFF}. Having derived the universal color and kinematic structure of these interactions,   the recursion relations between the components of the splitting functions can be cast in the form of a matrix equation, which has a particularly simple triangular form because of their causal structure. We have written a Mathematica code to solve  this equation to any desired order in opacity.

Even though the number of terms in the splitting kernels grows rapidly at higher orders in opacity, we can evaluate
these expressions numerically. One should note that the convergence of the VEGAS algorithm that approximates the probability distribution function of the integrand becomes slower and the calculation of the second order corrections uses  10 times larger number of random samples than that used in the leading order calculation.
The one-dimensional  parton splitting distribution  $\frac{dN}{dx}$ is shown in the right panel of Figure~\ref{f:ReactFF} for a $100 \; \mathrm{GeV}$ jet.   For light partons the corrections from second order in opacity in the small $x$ and large $x$  regions are all negative  and of order $30\% - 70\%$.  In the region of $x \sim 0.5$ the second order correction is positive and  boosts the branching probability.  This can also be clearly seen in the
insets of  the figure.  For heavy quarks 
2$^{\rm nd}$ order in opacity corrections are noticeably smaller in most parts of the available phase space.

\section{Conclusions}

In summary, I covered a range of subjects related to the theory and phenomenology of heavy flavor jets in heavy ion reactions - from the formal aspects of parton shower formation to the practical issue of identifying new observables  with better sensitivity to jet quenching effects that can enhance the existing tool chest of the field. Heavy flavor jets studies represent the convergence of the now well-developed heavy meson and light  jet theory and measurements. I expect that their role as precision diagnostics of nuclear matter will continue to grow in the future, especially  as they are being backed up by new theoretical  developments.

%\bibliographystyle{JHEP}
%\bibliography{PoS}

\providecommand{\href}[2]{#2}\begingroup\raggedright\endgroup

%\end{thebibliography}

\end{document}